\documentclass[aps, prl, twocolumn, superscriptaddress, showpacs, amssymb, amsmath]{revtex4-1}
\usepackage{graphicx}
\usepackage{hyperref}
\usepackage{float}
\usepackage{epstopdf}
\usepackage{color}
\usepackage{braket}

\begin{document}

\title{Topology in the Sierpi\'{n}ski-Hofstadter problem}
\date{\today}

\author{
Marta~Brzezi\'{n}ska}
\affiliation{
 Department of Theoretical Physics, Faculty of Fundamental Problems of Technology, Wroc\l{}aw University of Science and Technology, 50-370 Wroc\l{}aw, Poland}
 
\author{
Ashley~M.~Cook}
\affiliation{
 Department of Physics, University of Zurich, Winterthurerstrasse 190, 8057 Zurich, Switzerland
}

\author{
Titus~Neupert}
\affiliation{
 Department of Physics, University of Zurich, Winterthurerstrasse 190, 8057 Zurich, Switzerland
}

\begin{abstract}
Using the Sierpi\'{n}ski carpet and gasket, we investigate whether fractal lattices embedded in two-dimensional space can support topological phases when subjected to a homogeneous external magnetic field. To this end, we study the localization property of eigenstates, the Chern number, and the evolution of energy level statistics when disorder is introduced. Combining these theoretical tools, we identify regions in the phase diagram of both the carpet and the gasket, for which the systems exhibit properties normally associated to gapless topological phases with a mobility edge.
\end{abstract}
\maketitle

\textit{Introduction} ---
The topological character of electronic states of quantum matter is imprinted in
important universal characteristics, such as quantized response functions, localization properties of eigenstates, and protected boundary modes.
Whether a system can in principle support topological phases depends on its dimensionality and the set of symmetries with respect to which topology is defined. For noninteracting but potentially disordered systems, this information is tabulated in the ten-fold way~\cite{10fold}, while the classification of symmetry protected topological (SPT) phases in general also encompasses interactions~\cite{WenInteract}.

All classifications of topological states have so far been performed for systems of integer spatial dimension. For instance, the integer quantum Hall effect, which can in many ways be viewed as the most robust and fundamental topological phase, exists in two-dimensional systems, but not in one-dimensional systems (in three dimensions it can exist only as a weak phase whose essential properties are inherited from the two-dimensional realization~\cite{10fold, PhysRevLett.93.206602}). 
However, to define the topology of quantum states, only a notion of locality and the possibility to take a thermodynamic limit are required, both of which can be defined for a general graph, not only for a regular lattice. In particular, these concepts can be defined for a fractal lattice.
Thus, a notion of topological states should also exist for quantum states defined on general graphs, including fractals with (non-integer) Hausdorff dimension.
It is imperative to ask whether the quantum states on such graphs can in fact be topological and how the classification depends on properties of the fractals like their dimensionality or ramification number. 

Here, we investigate these questions by means of a case study on what might be considered the most natural candidate for a topological phase of a fractal lattice: the electronic structure in presence of a homogeneous magnetic field. Specifically, we study lattice regularizations of the Sierpi\'{n}ski carpet (SC) and the Sierpi\'{n}ski gasket (SG), i.e., the Sierpi\'{n}ski-Hofstadter problem. By considering a magnetic flux that is homogeneous with respect to the two-dimensional plane in which the fractals are embedded, we study the situation most relevant to meso- and nanoscopic experiments. An important difference between the gasket and the carpet is that their ramification number is finite and infinite, respectively. This means that an extensive part of a gasket can be separated by just cutting a finite number of bonds, while for the carpet this operation requires cutting a number of bonds that tends to infinity in the thermodynamic limit.

Renewed interest in the physics of fractals has been ignited by progress in experimental methods which allow to create fractal structures using, for example, molecular chains~\cite{Shang2015}, atomic manipulation of molecules on the surface~\cite{TriangleCu} or focused ion beam lithography~\cite{FIBSC}. Recent theoretical developments include quantum transport calculations~\cite{2016:TomadinTransport1}, investigations of optoelectronic properties~\cite{2017:YuanOptCond, plasmons}, random fractal lattices~\cite{2017:SachaRandFractal}, entanglement entropy and entanglement spectra in fractals~\cite{2015:MatsuedaSVD}, and systems with fractal boundaries~\cite{2014:LiTIFractal} or fractal-like structures hosting flat bands~\cite{flatbands}. The spectra of the SC and SG in a presence of a magnetic field were studied as well~\cite{GHEZ19871291,PhysRevB.29.5504, PhysRevLett.49.1194, PhysRevB.60.10054}, but possible topological properties of the eigenstates have not been investigated. Topological phases on fractal lattices have been examined only recently within Bernevig-Hughes-Zhang (BHZ) model on the SC and SG in Ref.~\onlinecite{2018:BHZ} (similar considerations for completely random lattices are presented in Ref.~\onlinecite{randomTI1}).

We use a combination of approaches to identify the topological properties of the Sierpi\'{n}ski-Hofstadter problem as a function of filling and magnetic flux. First, we analyze the localization properties of individual eigenstates on the lattice. Thereby we uncover a hierarchy of states sharply localized around ``holes" of the lattice. We identify the regions in the phase diagram where such states dominate. Second, we use a real-space formulation to compute the Chern number (or Hall conductivity). We find it to be sharply quantized to trivial and non-trivial values in parts of parameter space. Finally, we add disorder to the system and study the energy level spacing statistics. This way, we can identify regions in the phase diagram which are separated by a plateau transition from an Anderson insulating limit, indicating their non-trivial topology. We find good agreement between these regions and the ones with non-zero Chern number, confirming the consistency of our results. In the following, we present each of these three approaches in succession. Further details and a cross-check of our methods for the known Hofstadter problem on two-dimensional lattices are contained in the Supplemental Material.

\begin{figure}[t]
\centering
\includegraphics[width=0.95\columnwidth]{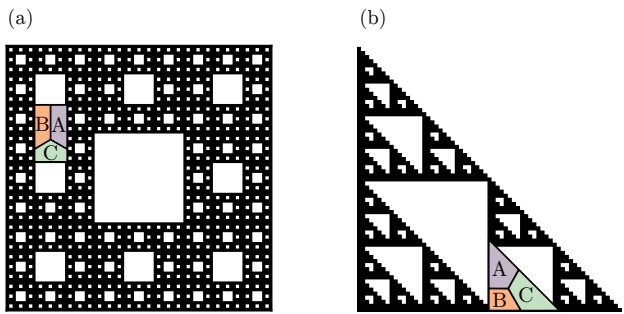}
\caption{Sierpi\'{n}ski (a) carpet and (b) gasket at iteration $n = 4$ and $n=6$, respectively. Black squares correspond to kept sites from underlying square (in case of SC) and triangular (SG) lattices. The summation regions included in real-space Chern number calculations are marked with A, B, C.} 
\label{fig:lattice}
\end{figure}

\textit{Model} ---
We consider a tight-binding Hamiltonian describing spinless fermions in a perpendicular orbital magnetic field 
\begin{equation}
H = -t  \sum_{\langle i, j \rangle} e^{\textnormal{i} A_{ij}} c^{\dagger}_i c_j + \mathrm{h.c.},
\label{eq:hamiltonian}
\end{equation}
where $c^{\dagger}_i (c_i)$ is a creation (annihilation) operator on lattice site $i$ and $\langle \ldots \rangle$ denotes nearest neighbors. The hopping integral $t$ is the only energy scale, and we set it to $t = 1$. The magnetic field is incorporated into the model by the phase factors $A_{ij} = \int_i^j \mathbf{A} \cdot d \mathbf{r}$ with $\mathbf{A}$ being the vector potential satisfying relation $\mathbf{B} =  \nabla \times \mathbf{A}$. Hamiltonian~\eqref{eq:hamiltonian} is studied on a graph that corresponds to the lattice-regulated SC and SG (see Fig.~\ref{fig:lattice}). These graphs have a smallest square (for the SC) and a smallest triangle (for the SG), respectively. The magnetic flux per this smallest element is chosen to be $\Phi$, a fraction $\alpha$ of the flux quantum $\Phi_0$ (where $\Phi_0 = 2 \pi$ in units where $\hbar = e = 1$), i.e., $\Phi  /  \Phi_0 = \alpha$. The magnetic field is assumed to be homogeneous in the two-dimensional space in which the fractal is embedded. To study the effect of disorder, we add the on-site disorder term $\sum_iV_i c_i^{\dagger} c_i$ to the Hamiltonian. The coefficients $V_i$ are randomly chosen values from the uniform distribution in the range $\left[ - \frac{W}{2}, \frac{W}{2} \right]$ where $W$ is the disorder strength in units of $t$.

\begin{figure*}[t]
\centering
\includegraphics[width=0.85\textwidth]{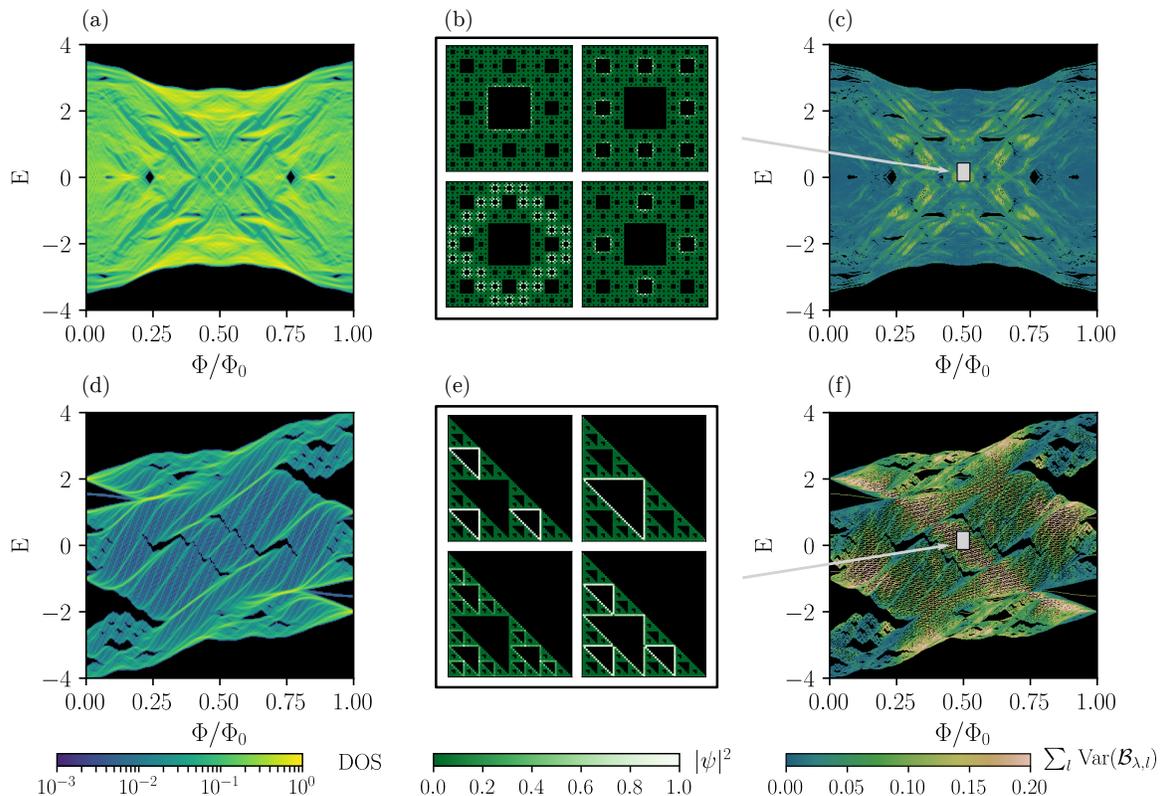}
\caption{(a, d) Density of states in the energy-flux plane, (b, e) localization of the eigenstates, and (c, f) edge-locality marker for the SC at iteration $n = 4$ and the SG at iteration $n = 6$, respectively. Darker regions are related to smaller density. Energy spectra reveal two gaps at $E =0$ for SC and numerous gaps in case of SG. Electronic densities presented in (b, e) correspond to the time-reversal symmetric point ($\alpha = 1/2$) indicated by a framed part of the spectrum. The color scale corresponds to the square modulus $| \psi_i|^2$ of the wave function normalized by its maximum value. (c, f) shows how the $\mathcal{B}_{\lambda, l}$ marker changes between consecutive eigenstates at fixed flux. Low DOS regions are associated with largely varying localization properties of the eigenstates.}
\label{fig:DOS_localiz}
\end{figure*}

\textit{Spectral and eigenstate localization properties} ---
In Fig.~\ref{fig:DOS_localiz} (a, d), we show the density of states (DOS) for systems with open boundary conditions as a function of $\alpha$. Discrete energy spectra $E_\lambda$ are smoothed using a Gaussian function $f(E, \alpha) = \sum_{\lambda} \exp \hspace*{0.1cm} \left\{ - \left[ E - E_{\lambda} (\alpha) \right]^2 / \eta \right\}$ with broadening $\eta = 0.001$. Similar to the Hofstadter problem on regular lattices, a presence of magnetic field gives rise to the self-similar structure of spectra in the energy-flux plane known as Hofstadter's butterfly~\cite{1976:Hofstadter}. The spectrum of the SC [Fig.~\ref{fig:DOS_localiz}~(a)] is reflection-symmetric both with respect to the $E = 0$ and the $\alpha = 1/2$ lines due to a chiral symmetry of this Hamiltonian on a bipartite lattice. A finite gapped region of maximal extend in energy $\sim 0.1$ is observed for small range of the flux around $\alpha = 1/4$ and $E = 0$. Regions of low DOS (appearing in dark blue color) host states with distinct localization properties, which we confirm below. The spectrum of the SG [Fig.~\ref{fig:DOS_localiz}~(d)] has only a point inversion symmetry about $\alpha=1/2$ and $E=0$, while reflection symmetries are lost for this non-bipartite lattice. At zero flux, the spectrum is known to be a fractal with discrete eigenvalues~\cite{PhysRevB.28.3110}. The magnetic field lifts these degeneracies. The most distinct spectral features are a large DOS at $\alpha = 1/4$, $E \approx 1.4$ as well as various fully gapped regions. In Fig.~\ref{fig:DOS_localiz}~(b, e) we present electronic densities for representative states at the time-reversal symmetric point ($\alpha =1/2$) slightly above zero energy. They reveal states sharply localized at the internal edges of the fractal at different levels of the hierarchy. Groups of states of this type can be found in very close spectral proximity to one another in various places of the phase diagram.
To map out these regions, we calculate a localization marker defined as
\begin{equation}
\mathcal{B}_{\lambda, l} = \sum_{i \in \mathcal{E}_l} | \psi_{\lambda,i} |^2, 
\end{equation}
where $\braket{i | \psi_{\lambda}} = \psi_{\lambda,i} $ and the summation is taken over the edges $\mathcal{E}_l$ of all internal triangles or squares at level $l$ of the hierarchy. Therefore, $\mathcal{B}_{\lambda, l}$ measures how much an eigenstate $\ket{\psi_{\lambda}}$ with an energy $E_{\lambda}$ is localized on the different edges of hierarchy level $l$. A similar hierarchy of edge-localized states was also observed for BHZ model in Ref.~\onlinecite{2018:BHZ}. With every $\ket{\psi_{\lambda}}$ we associate a set of $\mathcal{B}_{\lambda, l}$ for $l=0,\cdots, n$. To determine where in the phase diagram the localization properties are most rapidly varying, we calculate the variance for each entry of the set $\mathcal{B}_{\lambda, l}$, $l=0,\cdots, n$, across three consecutive states in the spectrum with energies $E_{\lambda - 1}$, $E_{\lambda}$ and $E_{\lambda+1}$ and sum these variances over $l$. The results are shown in Fig.~\ref{fig:DOS_localiz}~(c, f) and demonstrate that sharp changes in eigenstate localization appear predominantly in the regions with low DOS both for the carpet and the gasket. We are thus led to interpret the regions of low DOS as made of states with edge character at various levels of the fractal hierarchy.

\begin{figure*}[t]
\centering
\includegraphics[width=\textwidth]{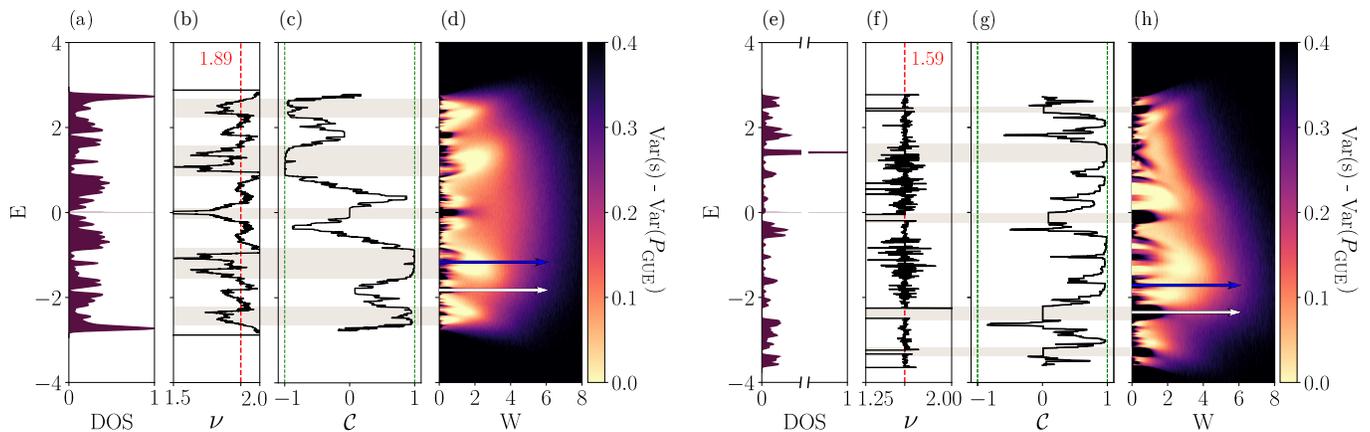}
\caption{(a, e) Density of states, (b, f) scaling exponent $\nu$ of the DOS with system size as a function of $E$, (c, g) the Chern number as a function of $E$ and (d, h) variance of level spacings in the energy-disorder plane at fixed flux $\alpha = 1/4$ for the SC and the SG, respectively. Grey rectangles are drawn to guide the eye. We identify spectral gaps to be topologically trivial for both lattices. Regions with quantized values of the Chern number close to 1 are separated by a delocalized state from the Anderson insulator limit, as it is the case for quantum Hall states  [blue arrows in (d, h)]. In contrast, a direct tranition to a fully localized phase is observed for states carrying trival Chern number as a function of $W$ [white arrows in (d, h)]. States with $\mathcal{C} \neq 0$ are characterized by a DOS scaling exponent $\nu$ smaller than $d_{\mathrm{H}}$ in (b). A deviation from that behavior is caused by singular peaks of the DOS.}
\label{fig:Chern_disord}
\end{figure*}

\textit{Real-space Chern number} ---
To study potential topological properties of the Sierpi\'{n}ski-Hofstadter problem, we adopt a real-space method to compute the Chern number introduced in Ref.~\onlinecite{kitaev}
\begin{equation}
\mathcal{C} = 12 \pi i \sum_{j \in A} \sum_{k \in B} \sum_{l \in C} \left( P_{jk} P_{kl} P_{lj} - P_{jl} P_{lk} P_{kj} \right),
\end{equation}
where $P$ is a projector onto occupied states with respect to a given Fermi level $E$ and $j, k, l$ are site indices in three distinct neighboring regions $A$, $B$, and $C$ of the lattice. The regions are three neighboring sectors arranged counterclockwise as shown in Fig.~\ref{fig:lattice}. If $\mathcal{C}$ is quantized, it becomes independent of the detailed choice of $A$, $B$, $C$ in the limit of where the number of sites in each sector tends to infinity.
We repeated the calculation for various choices of $A$, $B$, $C$ and found that the intervals of quantized $\mathcal{C}$ discussed below are robust.
In Fig. \ref{fig:Chern_disord}, we show $\mathcal{C}$ as a function of the Fermi energy $E$ at fixed value of flux $ \alpha =1/4$ for the $n=4$ iteration of SC (c) and the $n=6$ iteration of the SG (g). We obtain the following results: (i) All fully gapped regions of the spectrum, both in the case of SG and SC, carry $\mathcal{C}=0$. (ii) The regions of low but non-zero density of states (blue) in Fig.~\ref{fig:DOS_localiz}~(a) for the SC correspond to stable plateaus with $\mathcal{C} \sim \pm 1.0$ (for a wide range of energies $E = -1.5 \ldots - 0.9$ and $E = 0.9 \ldots 1.5$), together with less quantized regions with $\mathcal{C} \sim \pm 0.96$ ($E = -2.6\ldots -2.5$ and $E = 2.5\ldots 2.6$). Deviations from quantized Chern numbers are observed when the DOS is enhanced, for example around $E = -1.2$ and $E = 1.2$. (iii) For the SG, non-trivial regions are less clearly identifiable, but a clear plateau from $E = 1\ldots 1.6$ converges to $\mathcal{C} \sim 1.0$. 

To further substantiate the connection between the DOS and the Chern number, we calculate the number of states at fixed $\alpha$ averaged over an energy interval $[\epsilon - \delta,\epsilon + \delta]$ (with $\delta = 0.1$ for the SC and $\delta = 0.05$ for the SG) for different system sizes, and compute the average scaling exponent $\nu$ of the number of states in that energy range with system size. 
On average, $\nu$ equals the Hausdorff dimension $d_{\mathrm{H}}$. We show in Fig.~\ref{fig:Chern_disord}~(b) that for the SC regions with (nearly) quantized Chern number consistently show scaling with $\nu < d_{\mathrm{H}}$. This indicates that the normalized DOS would scale to zero in the thermodynamic limit in regions with quantized Chern number. 
For the SG, the situation is less clear except in regions of trivial Chern number where no states are found.
[see Fig.~\ref{fig:Chern_disord} (f)].

\textit{Level spacings analysis} ---
A complementary probe of topology can be obtained by studying the effect of localization by disorder. At large disorder strength (much larger than the band width), all states of a system become Anderson localized. However, if the system is in an insulating state with non-trivial Hall conductivity for small disorder, the transition to the Anderson insulator happens via a critical delocalized state at intermediate disorder values~\cite{PhysRevLett.105.115501}.
To probe this transition -- if present -- we study the energy level spacing statistics of the Sierpi\'{n}ski-Hofstadter problem in the presence of disorder.

To determine whether states are extended or localized, we perform an energy level statistics analysis. For a given energy $\epsilon$ and disorder realization $\{V_i\}$, we find two closest eigenvalues satisfying $ E_{\lambda, \{V_i\}} < \epsilon < E_{\lambda + 1, \{V_i\}}$, then calculate level spacings $s_{\epsilon, m, \{V_i\}} = E_{\lambda + m + 1, \{V_i\}} - E_{\lambda + m, \{V_i\}}$, where $m \in \lbrace -k, k \rbrace$, and normalize them. We set $k=2$ as suggested in Ref.~\onlinecite{2011:Prodan}. This allows to investigate the distribution of the level spacings and the variance $\mathrm{Var} (s_{\epsilon}) = \langle s_{\epsilon}^2 \rangle - \langle s_{\epsilon} \rangle^2$. The average is taken with respect $m$ and $10^3$ disorder realizations for fixed $\epsilon$. If states are delocalized, then the level spacings should obey the Wigner-Dyson surmise in the unitary case given by $P_{\mathrm{GUE}} (s) = \frac{32 s^2}{\pi^2} e^{-\frac{4}{\pi} s^2}$; if localized, they are expected to follow a Poisson distribution $P (s) = \exp (-s)$. Using the numerically obtained distribution of the level spacings for different disorder amplitudes $W$, we calculate the difference between $\mathrm{Var}(s)$ and the variance corresponding to $P_{\mathrm{GUE}}$ [see Fig.~\ref{fig:Chern_disord} (d, h)]. Since disorder calculations require exact diagonalization of the Hamiltonian repeatedly, we focus on smaller systems (iteration $n=3$ for SC and $n=5$ for SG). 
We find that regions in energy for which the Chern number is quantized consistently show a large $\mathrm{Var}(s)$ for small $W$, i.e., they are localized [see Fig.~\ref{fig:Chern_disord}~(d, h)]. 
At strong disorder the systems are fully localized as well. 
As one follows a line of increasing $W$ at constant energy, two transition scenarios can be found, corresponding to the white and blue arrows in Fig.~\ref{fig:Chern_disord}~(d, h), respectively: either there is a crossover into the localized region at large $W$ without $\mathrm{Var}(s)$ ever becoming close to $\mathrm{Var}(P_{\mathrm{GUE}}) = 0.178$, or a localized region at small $W$ is separated by a delocalized region with $\mathrm{Var}(P_{\mathrm{GUE}}) = 0.178$ from the localized states at large $W$. These two scenarios are in correspondence with the Chern numbers computed in the absence of disorder: The former is found for regions with trivial quantized Chern number, the latter for non-trivial quantized Chern number.

\textit{Conclusions} ---
We have investigated topological electronic properties of two fractal lattices, the Sierpi\'{n}ski carpet and gasket in a external magnetic field. By performing level spacings analysis, Chern number calculations and by investigating localization properties of individual eigenstates, we identified states with non-trivial topology that show characteristics similar to the quantum Hall effect. They do, however, occur on graphs with non-integer Hausdorff dimension. Our results, which strongly suggest the existence of quantum Hall-type states on fractals, call for an extension of the classification of topological states to such more general graphs. 

The example we investigated is in particular tailored to challenge the following sharp distinction by dimensionality: 
Long-range entangled phases (to which the integer quantum Hall effect belongs in the terminology of Ref.~\cite{WenClass1}) have been proven to not exist in one dimension~\cite{WenClass2}. It is thus imperative to ask what dimensional properties a graph must have in order to support long-range entangled ground states of local Hamiltonians.  

Finally, we emphasize that topological states on fractal lattices may provide a way to understand so-called fracton topological order in three-dimensional systems~\cite{Fracton1, Fracton2, Fracton3, Fracton4}. Fracton models have been studied as codes with large ground state manifolds in which quantum information may be stored. In some fracton models, operators that create excitations have support on a fractal subset of the three-dimensional lattice. From more conventional topological orders, we know that the operators that create excitations can be thought of carrying a lower-dimensional SPT phase. It would be interesting to investigate whether the same picture holds for the fractal case. Similar considerations may apply to the related fractal symmetry breaking states investigated in Ref.~\cite{SondhiFracSym}.

\textit{Acknowledgements} --- The authors thank Frank Schindler for discussions at the early stage of this work and in particular Shivaji Sondhi as well as Curt von Keyserlingk for discussions from which the question about topology in fractal lattices emerged. MB was supported by the Polish NCN Grant No. 2014/14/A/ST3/00654. TN acknowledges support by the Swiss National Science Foundation (grant number: 200021\_169061).

\bibliography{bibliography.bib}

\begin{thebibliography}{31}%
\makeatletter
\providecommand \@ifxundefined [1]{%
 \@ifx{#1\undefined}
}%
\providecommand \@ifnum [1]{%
 \ifnum #1\expandafter \@firstoftwo
 \else \expandafter \@secondoftwo
 \fi
}%
\providecommand \@ifx [1]{%
 \ifx #1\expandafter \@firstoftwo
 \else \expandafter \@secondoftwo
 \fi
}%
\providecommand \natexlab [1]{#1}%
\providecommand \enquote  [1]{``#1''}%
\providecommand \bibnamefont  [1]{#1}%
\providecommand \bibfnamefont [1]{#1}%
\providecommand \citenamefont [1]{#1}%
\providecommand \href@noop [0]{\@secondoftwo}%
\providecommand \href [0]{\begingroup \@sanitize@url \@href}%
\providecommand \@href[1]{\@@startlink{#1}\@@href}%
\providecommand \@@href[1]{\endgroup#1\@@endlink}%
\providecommand \@sanitize@url [0]{\catcode `\\12\catcode `\$12\catcode
  `\&12\catcode `\#12\catcode `\^12\catcode `\_12\catcode `\%12\relax}%
\providecommand \@@startlink[1]{}%
\providecommand \@@endlink[0]{}%
\providecommand \url  [0]{\begingroup\@sanitize@url \@url }%
\providecommand \@url [1]{\endgroup\@href {#1}{\urlprefix }}%
\providecommand \urlprefix  [0]{URL }%
\providecommand \Eprint [0]{\href }%
\providecommand \doibase [0]{http://dx.doi.org/}%
\providecommand \selectlanguage [0]{\@gobble}%
\providecommand \bibinfo  [0]{\@secondoftwo}%
\providecommand \bibfield  [0]{\@secondoftwo}%
\providecommand \translation [1]{[#1]}%
\providecommand \BibitemOpen [0]{}%
\providecommand \bibitemStop [0]{}%
\providecommand \bibitemNoStop [0]{.\EOS\space}%
\providecommand \EOS [0]{\spacefactor3000\relax}%
\providecommand \BibitemShut  [1]{\csname bibitem#1\endcsname}%
\let\auto@bib@innerbib\@empty
%</preamble>
\bibitem [{\citenamefont {Ryu}\ \emph {et~al.}(2010)\citenamefont {Ryu},
  \citenamefont {Schnyder}, \citenamefont {Furusaki},\ and\ \citenamefont
  {Ludwig}}]{10fold}%
  \BibitemOpen
  \bibfield  {author} {\bibinfo {author} {\bibfnamefont {S.}~\bibnamefont
  {Ryu}}, \bibinfo {author} {\bibfnamefont {A.~P.}\ \bibnamefont {Schnyder}},
  \bibinfo {author} {\bibfnamefont {A.}~\bibnamefont {Furusaki}}, \ and\
  \bibinfo {author} {\bibfnamefont {A.~W.~W.}\ \bibnamefont {Ludwig}},\ }\href
  {http://stacks.iop.org/1367-2630/12/i=6/a=065010} {\bibfield  {journal}
  {\bibinfo  {journal} {New Journal of Physics}\ }\textbf {\bibinfo {volume}
  {12}},\ \bibinfo {pages} {065010} (\bibinfo {year} {2010})}\BibitemShut
  {NoStop}%
\bibitem [{\citenamefont {Gu}\ and\ \citenamefont {Wen}(2014)}]{WenInteract}%
  \BibitemOpen
  \bibfield  {author} {\bibinfo {author} {\bibfnamefont {Z.-C.}\ \bibnamefont
  {Gu}}\ and\ \bibinfo {author} {\bibfnamefont {X.-G.}\ \bibnamefont {Wen}},\
  }\href {\doibase 10.1103/PhysRevB.90.115141} {\bibfield  {journal} {\bibinfo
  {journal} {Phys. Rev. B}\ }\textbf {\bibinfo {volume} {90}},\ \bibinfo
  {pages} {115141} (\bibinfo {year} {2014})}\BibitemShut {NoStop}%
\bibitem [{\citenamefont {Haldane}(2004)}]{PhysRevLett.93.206602}%
  \BibitemOpen
  \bibfield  {author} {\bibinfo {author} {\bibfnamefont {F.~D.~M.}\
  \bibnamefont {Haldane}},\ }\href {\doibase 10.1103/PhysRevLett.93.206602}
  {\bibfield  {journal} {\bibinfo  {journal} {Phys. Rev. Lett.}\ }\textbf
  {\bibinfo {volume} {93}},\ \bibinfo {pages} {206602} (\bibinfo {year}
  {2004})}\BibitemShut {NoStop}%
\bibitem [{\citenamefont {Shang}\ \emph {et~al.}(2015)\citenamefont {Shang},
  \citenamefont {Wang}, \citenamefont {Chen}, \citenamefont {Dai},
  \citenamefont {Zhou}, \citenamefont {Kuttner}, \citenamefont {Hilt},
  \citenamefont {Shao}, \citenamefont {Gottfried},\ and\ \citenamefont
  {Wu}}]{Shang2015}%
  \BibitemOpen
  \bibfield  {author} {\bibinfo {author} {\bibfnamefont {J.}~\bibnamefont
  {Shang}}, \bibinfo {author} {\bibfnamefont {Y.}~\bibnamefont {Wang}},
  \bibinfo {author} {\bibfnamefont {M.}~\bibnamefont {Chen}}, \bibinfo {author}
  {\bibfnamefont {J.}~\bibnamefont {Dai}}, \bibinfo {author} {\bibfnamefont
  {X.}~\bibnamefont {Zhou}}, \bibinfo {author} {\bibfnamefont {J.}~\bibnamefont
  {Kuttner}}, \bibinfo {author} {\bibfnamefont {G.}~\bibnamefont {Hilt}},
  \bibinfo {author} {\bibfnamefont {X.}~\bibnamefont {Shao}}, \bibinfo {author}
  {\bibfnamefont {J.~M.}\ \bibnamefont {Gottfried}}, \ and\ \bibinfo {author}
  {\bibfnamefont {K.}~\bibnamefont {Wu}},\ }\href
  {http://dx.doi.org/10.1038/nchem.2211} {\bibfield  {journal} {\bibinfo
  {journal} {Nature Chemistry}\ }\textbf {\bibinfo {volume} {7}},\ \bibinfo
  {pages} {389} (\bibinfo {year} {2015})}\BibitemShut {NoStop}%
\bibitem [{\citenamefont {{Kempkes}}\ \emph {et~al.}(2018)\citenamefont
  {{Kempkes}}, \citenamefont {{Slot}}, \citenamefont {{Freeney}}, \citenamefont
  {{Zevenhuizen}}, \citenamefont {{Vanmaekelbergh}}, \citenamefont {{Swart}},\
  and\ \citenamefont {{Morais Smith}}}]{TriangleCu}%
  \BibitemOpen
  \bibfield  {author} {\bibinfo {author} {\bibfnamefont {S.~N.}\ \bibnamefont
  {{Kempkes}}}, \bibinfo {author} {\bibfnamefont {M.~R.}\ \bibnamefont
  {{Slot}}}, \bibinfo {author} {\bibfnamefont {S.~E.}\ \bibnamefont
  {{Freeney}}}, \bibinfo {author} {\bibfnamefont {S.~J.~M.}\ \bibnamefont
  {{Zevenhuizen}}}, \bibinfo {author} {\bibfnamefont {D.}~\bibnamefont
  {{Vanmaekelbergh}}}, \bibinfo {author} {\bibfnamefont {I.}~\bibnamefont
  {{Swart}}}, \ and\ \bibinfo {author} {\bibfnamefont {C.}~\bibnamefont
  {{Morais Smith}}},\ }\href@noop {} {\  (\bibinfo {year} {2018})},\ \Eprint
  {http://arxiv.org/abs/1803.04698} {arXiv:1803.04698 [cond-mat.mes-hall]}
  \BibitemShut {NoStop}%
\bibitem [{\citenamefont {Chen}\ \emph {et~al.}(2014)\citenamefont {Chen},
  \citenamefont {Dikken}, \citenamefont {Prangsma}, \citenamefont {Segerink},\
  and\ \citenamefont {Herek}}]{FIBSC}%
  \BibitemOpen
  \bibfield  {author} {\bibinfo {author} {\bibfnamefont {T.~L.}\ \bibnamefont
  {Chen}}, \bibinfo {author} {\bibfnamefont {D.~J.}\ \bibnamefont {Dikken}},
  \bibinfo {author} {\bibfnamefont {J.~C.}\ \bibnamefont {Prangsma}}, \bibinfo
  {author} {\bibfnamefont {F.}~\bibnamefont {Segerink}}, \ and\ \bibinfo
  {author} {\bibfnamefont {J.~L.}\ \bibnamefont {Herek}},\ }\href
  {http://stacks.iop.org/1367-2630/16/i=9/a=093024} {\bibfield  {journal}
  {\bibinfo  {journal} {New Journal of Physics}\ }\textbf {\bibinfo {volume}
  {16}},\ \bibinfo {pages} {093024} (\bibinfo {year} {2014})}\BibitemShut
  {NoStop}%
\bibitem [{\citenamefont {van Veen}\ \emph {et~al.}(2016)\citenamefont {van
  Veen}, \citenamefont {Yuan}, \citenamefont {Katsnelson}, \citenamefont
  {Polini},\ and\ \citenamefont {Tomadin}}]{2016:TomadinTransport1}%
  \BibitemOpen
  \bibfield  {author} {\bibinfo {author} {\bibfnamefont {E.}~\bibnamefont {van
  Veen}}, \bibinfo {author} {\bibfnamefont {S.}~\bibnamefont {Yuan}}, \bibinfo
  {author} {\bibfnamefont {M.~I.}\ \bibnamefont {Katsnelson}}, \bibinfo
  {author} {\bibfnamefont {M.}~\bibnamefont {Polini}}, \ and\ \bibinfo {author}
  {\bibfnamefont {A.}~\bibnamefont {Tomadin}},\ }\href {\doibase
  10.1103/PhysRevB.93.115428} {\bibfield  {journal} {\bibinfo  {journal} {Phys.
  Rev. B}\ }\textbf {\bibinfo {volume} {93}},\ \bibinfo {pages} {115428}
  (\bibinfo {year} {2016})}\BibitemShut {NoStop}%
\bibitem [{\citenamefont {van Veen}\ \emph {et~al.}(2017)\citenamefont {van
  Veen}, \citenamefont {Tomadin}, \citenamefont {Polini}, \citenamefont
  {Katsnelson},\ and\ \citenamefont {Yuan}}]{2017:YuanOptCond}%
  \BibitemOpen
  \bibfield  {author} {\bibinfo {author} {\bibfnamefont {E.}~\bibnamefont {van
  Veen}}, \bibinfo {author} {\bibfnamefont {A.}~\bibnamefont {Tomadin}},
  \bibinfo {author} {\bibfnamefont {M.}~\bibnamefont {Polini}}, \bibinfo
  {author} {\bibfnamefont {M.~I.}\ \bibnamefont {Katsnelson}}, \ and\ \bibinfo
  {author} {\bibfnamefont {S.}~\bibnamefont {Yuan}},\ }\href {\doibase
  10.1103/PhysRevB.96.235438} {\bibfield  {journal} {\bibinfo  {journal} {Phys.
  Rev. B}\ }\textbf {\bibinfo {volume} {96}},\ \bibinfo {pages} {235438}
  (\bibinfo {year} {2017})}\BibitemShut {NoStop}%
\bibitem [{\citenamefont {Westerhout}\ \emph {et~al.}(2018)\citenamefont
  {Westerhout}, \citenamefont {van Veen}, \citenamefont {Katsnelson},\ and\
  \citenamefont {Yuan}}]{plasmons}%
  \BibitemOpen
  \bibfield  {author} {\bibinfo {author} {\bibfnamefont {T.}~\bibnamefont
  {Westerhout}}, \bibinfo {author} {\bibfnamefont {E.}~\bibnamefont {van
  Veen}}, \bibinfo {author} {\bibfnamefont {M.~I.}\ \bibnamefont {Katsnelson}},
  \ and\ \bibinfo {author} {\bibfnamefont {S.}~\bibnamefont {Yuan}},\ }\href
  {\doibase 10.1103/PhysRevB.97.205434} {\bibfield  {journal} {\bibinfo
  {journal} {Phys. Rev. B}\ }\textbf {\bibinfo {volume} {97}},\ \bibinfo
  {pages} {205434} (\bibinfo {year} {2018})}\BibitemShut {NoStop}%
\bibitem [{\citenamefont {Kosior}\ and\ \citenamefont
  {Sacha}(2017)}]{2017:SachaRandFractal}%
  \BibitemOpen
  \bibfield  {author} {\bibinfo {author} {\bibfnamefont {A.}~\bibnamefont
  {Kosior}}\ and\ \bibinfo {author} {\bibfnamefont {K.}~\bibnamefont {Sacha}},\
  }\href {\doibase 10.1103/PhysRevB.95.104206} {\bibfield  {journal} {\bibinfo
  {journal} {Phys. Rev. B}\ }\textbf {\bibinfo {volume} {95}},\ \bibinfo
  {pages} {104206} (\bibinfo {year} {2017})}\BibitemShut {NoStop}%
\bibitem [{\citenamefont {Lee}\ \emph {et~al.}(2015)\citenamefont {Lee},
  \citenamefont {Yamada}, \citenamefont {Kumamoto},\ and\ \citenamefont
  {Matsueda}}]{2015:MatsuedaSVD}%
  \BibitemOpen
  \bibfield  {author} {\bibinfo {author} {\bibfnamefont {C.~H.}\ \bibnamefont
  {Lee}}, \bibinfo {author} {\bibfnamefont {Y.}~\bibnamefont {Yamada}},
  \bibinfo {author} {\bibfnamefont {T.}~\bibnamefont {Kumamoto}}, \ and\
  \bibinfo {author} {\bibfnamefont {H.}~\bibnamefont {Matsueda}},\ }\href
  {\doibase 10.7566/JPSJ.84.013001} {\bibfield  {journal} {\bibinfo  {journal}
  {Journal of the Physical Society of Japan}\ }\textbf {\bibinfo {volume}
  {84}},\ \bibinfo {pages} {013001} (\bibinfo {year} {2015})}\BibitemShut
  {NoStop}%
\bibitem [{\citenamefont {Song}\ \emph {et~al.}(2014)\citenamefont {Song},
  \citenamefont {Zhang},\ and\ \citenamefont {Li}}]{2014:LiTIFractal}%
  \BibitemOpen
  \bibfield  {author} {\bibinfo {author} {\bibfnamefont {Z.-G.}\ \bibnamefont
  {Song}}, \bibinfo {author} {\bibfnamefont {Y.-Y.}\ \bibnamefont {Zhang}}, \
  and\ \bibinfo {author} {\bibfnamefont {S.-S.}\ \bibnamefont {Li}},\ }\href
  {\doibase 10.1063/1.4882166} {\bibfield  {journal} {\bibinfo  {journal}
  {Applied Physics Letters}\ }\textbf {\bibinfo {volume} {104}},\ \bibinfo
  {pages} {233106} (\bibinfo {year} {2014})}\BibitemShut {NoStop}%
\bibitem [{\citenamefont {Pal}\ and\ \citenamefont {Saha}(2018)}]{flatbands}%
  \BibitemOpen
  \bibfield  {author} {\bibinfo {author} {\bibfnamefont {B.}~\bibnamefont
  {Pal}}\ and\ \bibinfo {author} {\bibfnamefont {K.}~\bibnamefont {Saha}},\
  }\href {\doibase 10.1103/PhysRevB.97.195101} {\bibfield  {journal} {\bibinfo
  {journal} {Phys. Rev. B}\ }\textbf {\bibinfo {volume} {97}},\ \bibinfo
  {pages} {195101} (\bibinfo {year} {2018})}\BibitemShut {NoStop}%
\bibitem [{\citenamefont {Ghez}\ \emph {et~al.}(1987)\citenamefont {Ghez},
  \citenamefont {Wang}, \citenamefont {Rammal}, \citenamefont {Pannetier},\
  and\ \citenamefont {Bellissard}}]{GHEZ19871291}%
  \BibitemOpen
  \bibfield  {author} {\bibinfo {author} {\bibfnamefont {J.}~\bibnamefont
  {Ghez}}, \bibinfo {author} {\bibfnamefont {Y.~Y.}\ \bibnamefont {Wang}},
  \bibinfo {author} {\bibfnamefont {R.}~\bibnamefont {Rammal}}, \bibinfo
  {author} {\bibfnamefont {B.}~\bibnamefont {Pannetier}}, \ and\ \bibinfo
  {author} {\bibfnamefont {J.}~\bibnamefont {Bellissard}},\ }\href {\doibase
  https://doi.org/10.1016/0038-1098(87)90628-4} {\bibfield  {journal} {\bibinfo
   {journal} {Solid State Communications}\ }\textbf {\bibinfo {volume} {64}},\
  \bibinfo {pages} {1291 } (\bibinfo {year} {1987})}\BibitemShut {NoStop}%
\bibitem [{\citenamefont {Alexander}(1984)}]{PhysRevB.29.5504}%
  \BibitemOpen
  \bibfield  {author} {\bibinfo {author} {\bibfnamefont {S.}~\bibnamefont
  {Alexander}},\ }\href {\doibase 10.1103/PhysRevB.29.5504} {\bibfield
  {journal} {\bibinfo  {journal} {Phys. Rev. B}\ }\textbf {\bibinfo {volume}
  {29}},\ \bibinfo {pages} {5504} (\bibinfo {year} {1984})}\BibitemShut
  {NoStop}%
\bibitem [{\citenamefont {Rammal}\ and\ \citenamefont
  {Toulouse}(1982)}]{PhysRevLett.49.1194}%
  \BibitemOpen
  \bibfield  {author} {\bibinfo {author} {\bibfnamefont {R.}~\bibnamefont
  {Rammal}}\ and\ \bibinfo {author} {\bibfnamefont {G.}~\bibnamefont
  {Toulouse}},\ }\href {\doibase 10.1103/PhysRevLett.49.1194} {\bibfield
  {journal} {\bibinfo  {journal} {Phys. Rev. Lett.}\ }\textbf {\bibinfo
  {volume} {49}},\ \bibinfo {pages} {1194} (\bibinfo {year}
  {1982})}\BibitemShut {NoStop}%
\bibitem [{\citenamefont {Lindquist}\ and\ \citenamefont
  {Riklund}(1999)}]{PhysRevB.60.10054}%
  \BibitemOpen
  \bibfield  {author} {\bibinfo {author} {\bibfnamefont {B.}~\bibnamefont
  {Lindquist}}\ and\ \bibinfo {author} {\bibfnamefont {R.}~\bibnamefont
  {Riklund}},\ }\href {\doibase 10.1103/PhysRevB.60.10054} {\bibfield
  {journal} {\bibinfo  {journal} {Phys. Rev. B}\ }\textbf {\bibinfo {volume}
  {60}},\ \bibinfo {pages} {10054} (\bibinfo {year} {1999})}\BibitemShut
  {NoStop}%
\bibitem [{\citenamefont {{Agarwala}}\ \emph {et~al.}(2018)\citenamefont
  {{Agarwala}}, \citenamefont {{Pai}},\ and\ \citenamefont
  {{Shenoy}}}]{2018:BHZ}%
  \BibitemOpen
  \bibfield  {author} {\bibinfo {author} {\bibfnamefont {A.}~\bibnamefont
  {{Agarwala}}}, \bibinfo {author} {\bibfnamefont {S.}~\bibnamefont {{Pai}}}, \
  and\ \bibinfo {author} {\bibfnamefont {V.~B.}\ \bibnamefont {{Shenoy}}},\
  }\href@noop {} {\bibfield  {journal} {\bibinfo  {journal} {ArXiv e-prints}\ }
  (\bibinfo {year} {2018})},\ \Eprint {http://arxiv.org/abs/1803.01404}
  {arXiv:1803.01404 [cond-mat.dis-nn]} \BibitemShut {NoStop}%
\bibitem [{\citenamefont {Agarwala}\ and\ \citenamefont
  {Shenoy}(2017)}]{randomTI1}%
  \BibitemOpen
  \bibfield  {author} {\bibinfo {author} {\bibfnamefont {A.}~\bibnamefont
  {Agarwala}}\ and\ \bibinfo {author} {\bibfnamefont {V.~B.}\ \bibnamefont
  {Shenoy}},\ }\href {\doibase 10.1103/PhysRevLett.118.236402} {\bibfield
  {journal} {\bibinfo  {journal} {Phys. Rev. Lett.}\ }\textbf {\bibinfo
  {volume} {118}},\ \bibinfo {pages} {236402} (\bibinfo {year}
  {2017})}\BibitemShut {NoStop}%
\bibitem [{\citenamefont {Hofstadter}(1976)}]{1976:Hofstadter}%
  \BibitemOpen
  \bibfield  {author} {\bibinfo {author} {\bibfnamefont {D.~R.}\ \bibnamefont
  {Hofstadter}},\ }\href {\doibase 10.1103/PhysRevB.14.2239} {\bibfield
  {journal} {\bibinfo  {journal} {Phys. Rev. B}\ }\textbf {\bibinfo {volume}
  {14}},\ \bibinfo {pages} {2239} (\bibinfo {year} {1976})}\BibitemShut
  {NoStop}%
\bibitem [{\citenamefont {Domany}\ \emph {et~al.}(1983)\citenamefont {Domany},
  \citenamefont {Alexander}, \citenamefont {Bensimon},\ and\ \citenamefont
  {Kadanoff}}]{PhysRevB.28.3110}%
  \BibitemOpen
  \bibfield  {author} {\bibinfo {author} {\bibfnamefont {E.}~\bibnamefont
  {Domany}}, \bibinfo {author} {\bibfnamefont {S.}~\bibnamefont {Alexander}},
  \bibinfo {author} {\bibfnamefont {D.}~\bibnamefont {Bensimon}}, \ and\
  \bibinfo {author} {\bibfnamefont {L.~P.}\ \bibnamefont {Kadanoff}},\ }\href
  {\doibase 10.1103/PhysRevB.28.3110} {\bibfield  {journal} {\bibinfo
  {journal} {Phys. Rev. B}\ }\textbf {\bibinfo {volume} {28}},\ \bibinfo
  {pages} {3110} (\bibinfo {year} {1983})}\BibitemShut {NoStop}%
\bibitem [{\citenamefont {Kitaev}(2006)}]{kitaev}%
  \BibitemOpen
  \bibfield  {author} {\bibinfo {author} {\bibfnamefont {A.}~\bibnamefont
  {Kitaev}},\ }\href {\doibase https://doi.org/10.1016/j.aop.2005.10.005}
  {\bibfield  {journal} {\bibinfo  {journal} {Annals of Physics}\ }\textbf
  {\bibinfo {volume} {321}},\ \bibinfo {pages} {2 } (\bibinfo {year} {2006})},\
  \bibinfo {note} {january Special Issue}\BibitemShut {NoStop}%
\bibitem [{\citenamefont {Prodan}\ \emph {et~al.}(2010)\citenamefont {Prodan},
  \citenamefont {Hughes},\ and\ \citenamefont
  {Bernevig}}]{PhysRevLett.105.115501}%
  \BibitemOpen
  \bibfield  {author} {\bibinfo {author} {\bibfnamefont {E.}~\bibnamefont
  {Prodan}}, \bibinfo {author} {\bibfnamefont {T.~L.}\ \bibnamefont {Hughes}},
  \ and\ \bibinfo {author} {\bibfnamefont {B.~A.}\ \bibnamefont {Bernevig}},\
  }\href {\doibase 10.1103/PhysRevLett.105.115501} {\bibfield  {journal}
  {\bibinfo  {journal} {Phys. Rev. Lett.}\ }\textbf {\bibinfo {volume} {105}},\
  \bibinfo {pages} {115501} (\bibinfo {year} {2010})}\BibitemShut {NoStop}%
\bibitem [{\citenamefont {Prodan}(2011)}]{2011:Prodan}%
  \BibitemOpen
  \bibfield  {author} {\bibinfo {author} {\bibfnamefont {E.}~\bibnamefont
  {Prodan}},\ }\href {http://stacks.iop.org/1751-8121/44/i=11/a=113001}
  {\bibfield  {journal} {\bibinfo  {journal} {Journal of Physics A:
  Mathematical and Theoretical}\ }\textbf {\bibinfo {volume} {44}},\ \bibinfo
  {pages} {113001} (\bibinfo {year} {2011})}\BibitemShut {NoStop}%
\bibitem [{\citenamefont {Chen}\ \emph {et~al.}(2010)\citenamefont {Chen},
  \citenamefont {Gu},\ and\ \citenamefont {Wen}}]{WenClass1}%
  \BibitemOpen
  \bibfield  {author} {\bibinfo {author} {\bibfnamefont {X.}~\bibnamefont
  {Chen}}, \bibinfo {author} {\bibfnamefont {Z.-C.}\ \bibnamefont {Gu}}, \ and\
  \bibinfo {author} {\bibfnamefont {X.-G.}\ \bibnamefont {Wen}},\ }\href
  {\doibase 10.1103/PhysRevB.82.155138} {\bibfield  {journal} {\bibinfo
  {journal} {Phys. Rev. B}\ }\textbf {\bibinfo {volume} {82}},\ \bibinfo
  {pages} {155138} (\bibinfo {year} {2010})}\BibitemShut {NoStop}%
\bibitem [{\citenamefont {Chen}\ \emph {et~al.}(2011)\citenamefont {Chen},
  \citenamefont {Gu},\ and\ \citenamefont {Wen}}]{WenClass2}%
  \BibitemOpen
  \bibfield  {author} {\bibinfo {author} {\bibfnamefont {X.}~\bibnamefont
  {Chen}}, \bibinfo {author} {\bibfnamefont {Z.-C.}\ \bibnamefont {Gu}}, \ and\
  \bibinfo {author} {\bibfnamefont {X.-G.}\ \bibnamefont {Wen}},\ }\href
  {\doibase 10.1103/PhysRevB.83.035107} {\bibfield  {journal} {\bibinfo
  {journal} {Phys. Rev. B}\ }\textbf {\bibinfo {volume} {83}},\ \bibinfo
  {pages} {035107} (\bibinfo {year} {2011})}\BibitemShut {NoStop}%
\bibitem [{\citenamefont {Chamon}(2005)}]{Fracton1}%
  \BibitemOpen
  \bibfield  {author} {\bibinfo {author} {\bibfnamefont {C.}~\bibnamefont
  {Chamon}},\ }\href {\doibase 10.1103/PhysRevLett.94.040402} {\bibfield
  {journal} {\bibinfo  {journal} {Phys. Rev. Lett.}\ }\textbf {\bibinfo
  {volume} {94}},\ \bibinfo {pages} {040402} (\bibinfo {year}
  {2005})}\BibitemShut {NoStop}%
\bibitem [{\citenamefont {Haah}(2011)}]{Fracton2}%
  \BibitemOpen
  \bibfield  {author} {\bibinfo {author} {\bibfnamefont {J.}~\bibnamefont
  {Haah}},\ }\href {\doibase 10.1103/PhysRevA.83.042330} {\bibfield  {journal}
  {\bibinfo  {journal} {Phys. Rev. A}\ }\textbf {\bibinfo {volume} {83}},\
  \bibinfo {pages} {042330} (\bibinfo {year} {2011})}\BibitemShut {NoStop}%
\bibitem [{\citenamefont {Vijay}\ \emph {et~al.}(2016)\citenamefont {Vijay},
  \citenamefont {Haah},\ and\ \citenamefont {Fu}}]{Fracton3}%
  \BibitemOpen
  \bibfield  {author} {\bibinfo {author} {\bibfnamefont {S.}~\bibnamefont
  {Vijay}}, \bibinfo {author} {\bibfnamefont {J.}~\bibnamefont {Haah}}, \ and\
  \bibinfo {author} {\bibfnamefont {L.}~\bibnamefont {Fu}},\ }\href {\doibase
  10.1103/PhysRevB.94.235157} {\bibfield  {journal} {\bibinfo  {journal} {Phys.
  Rev. B}\ }\textbf {\bibinfo {volume} {94}},\ \bibinfo {pages} {235157}
  (\bibinfo {year} {2016})}\BibitemShut {NoStop}%
\bibitem [{\citenamefont {Hal\'asz}\ \emph {et~al.}(2017)\citenamefont
  {Hal\'asz}, \citenamefont {Hsieh},\ and\ \citenamefont {Balents}}]{Fracton4}%
  \BibitemOpen
  \bibfield  {author} {\bibinfo {author} {\bibfnamefont {G.~B.}\ \bibnamefont
  {Hal\'asz}}, \bibinfo {author} {\bibfnamefont {T.~H.}\ \bibnamefont {Hsieh}},
  \ and\ \bibinfo {author} {\bibfnamefont {L.}~\bibnamefont {Balents}},\ }\href
  {\doibase 10.1103/PhysRevLett.119.257202} {\bibfield  {journal} {\bibinfo
  {journal} {Phys. Rev. Lett.}\ }\textbf {\bibinfo {volume} {119}},\ \bibinfo
  {pages} {257202} (\bibinfo {year} {2017})}\BibitemShut {NoStop}%
\bibitem [{\citenamefont {{Devakul}}\ \emph {et~al.}(2018)\citenamefont
  {{Devakul}}, \citenamefont {{You}}, \citenamefont {{Burnell}},\ and\
  \citenamefont {{Sondhi}}}]{SondhiFracSym}%
  \BibitemOpen
  \bibfield  {author} {\bibinfo {author} {\bibfnamefont {T.}~\bibnamefont
  {{Devakul}}}, \bibinfo {author} {\bibfnamefont {Y.}~\bibnamefont {{You}}},
  \bibinfo {author} {\bibfnamefont {F.~J.}\ \bibnamefont {{Burnell}}}, \ and\
  \bibinfo {author} {\bibfnamefont {S.~L.}\ \bibnamefont {{Sondhi}}},\
  }\href@noop {} {\  (\bibinfo {year} {2018})},\ \Eprint
  {http://arxiv.org/abs/1805.04097} {arXiv:1805.04097 [cond-mat.str-el]}
  \BibitemShut {NoStop}%
\end{thebibliography}%
\clearpage

\onecolumngrid
\begin{center}
\textbf{\large Supplemental Materials}
\end{center}
\appendix
\section{Creating the lattices and flux distribution}
\begin{figure}[H]
\centering
\includegraphics[width=0.5\columnwidth]{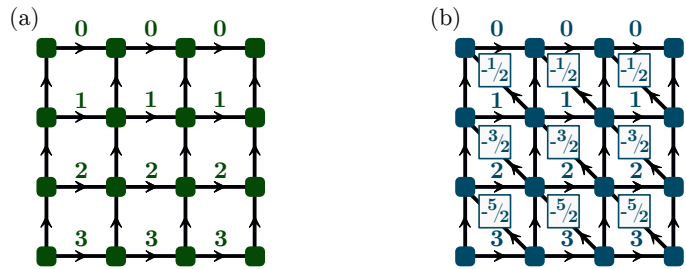}
\caption{Phase distribution on $4 \times 4$ (a) square and (b) triangle lattices with open boundary conditions. $A_{ij}$ phase between site $i$ and $j$ is equal to the number shown above the bond in $2 \pi$ units. A phase acquired with the respect to the direction pointed by arrows has a positive sign.}
\label{fig:FluxDistr}
\end{figure}

To construct the Sierpi\'{n}ski carpet (with $d_H \simeq 1.89$), we start from a square lattice with $L(n) = 3^n$ sites along outer edge and in every iterative construction step $n$ we remove $(1-(8/9)^n) \cdot 9^n$ sites. From Pascal's triangle modulo prime number $m$ embedded in triangular lattice with $2^n + 1$ rows, we can create a series of fractal triangular lattices with Hausdorff dimension $d_H = 1 +\log_m \left( \frac{m+ 1}{2} \right)$ ($m = 2$ gives the Sierpi\'{n}ski gasket with $d_H \simeq 1.59$). Fig. \ref{fig:FluxDistr} presents Peierls phase factors distribution. Magnetic flux through plaquette (being the smallest square or triangle in case of SC or SG, respectively) is equal to an integer multiple of $2\pi$.

\section{Level statistics}
\begin{figure}
\centering
\includegraphics[width=0.7\columnwidth]{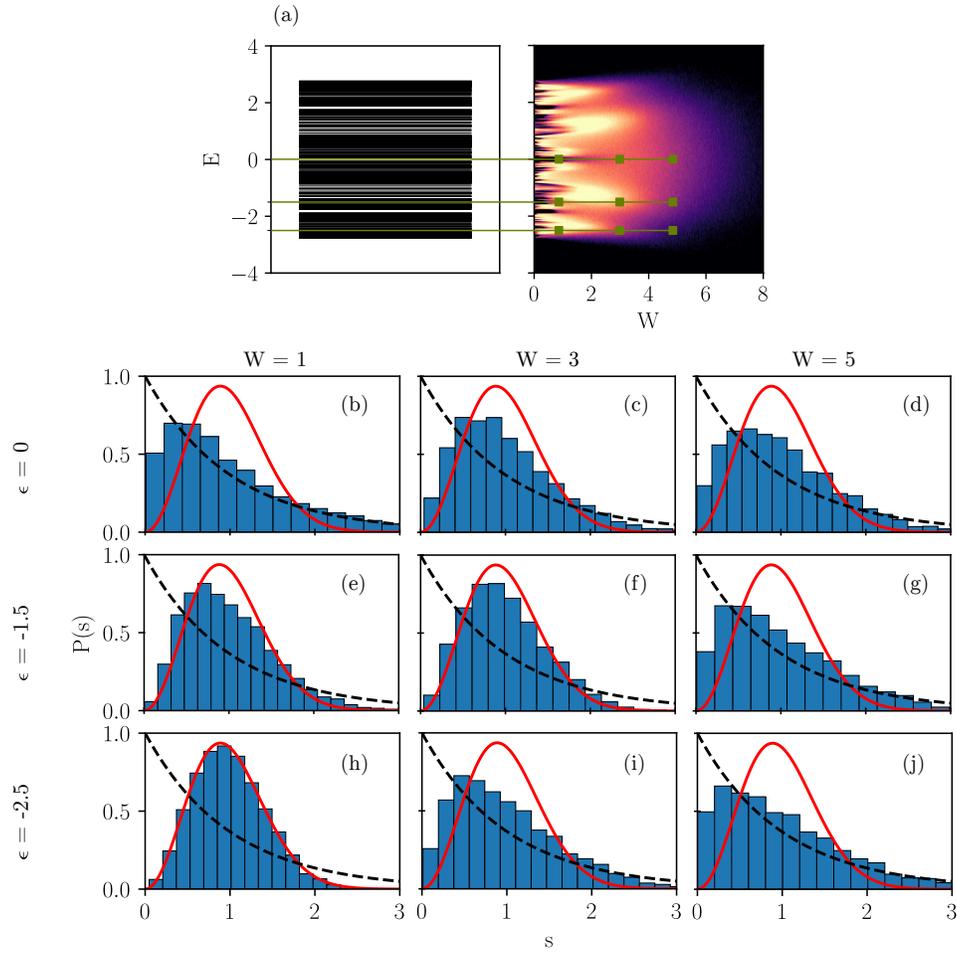}
\caption{(a) Energy levels for $n=3$ carpet in the absence of disorder and phase diagram from \ref{fig:Chern_disord} at $\alpha = 1/4$. (b)--(j) distribution of the level spacings. Histograms are related to the spacings around (h, i, j) $\epsilon = -2.5$, (e, f, g) - to $\epsilon = -1.5$ and (b, c, d) - to $\epsilon = 0$. Calculations were performed for three disorder strengths $W = 1$ (b, e, h), $W = 3$ (c, f, i) and $W = 5$ (d, g, j)}
\label{fig:hist_SC}
\end{figure}

\begin{figure}
\centering
\includegraphics[width=0.7\columnwidth]{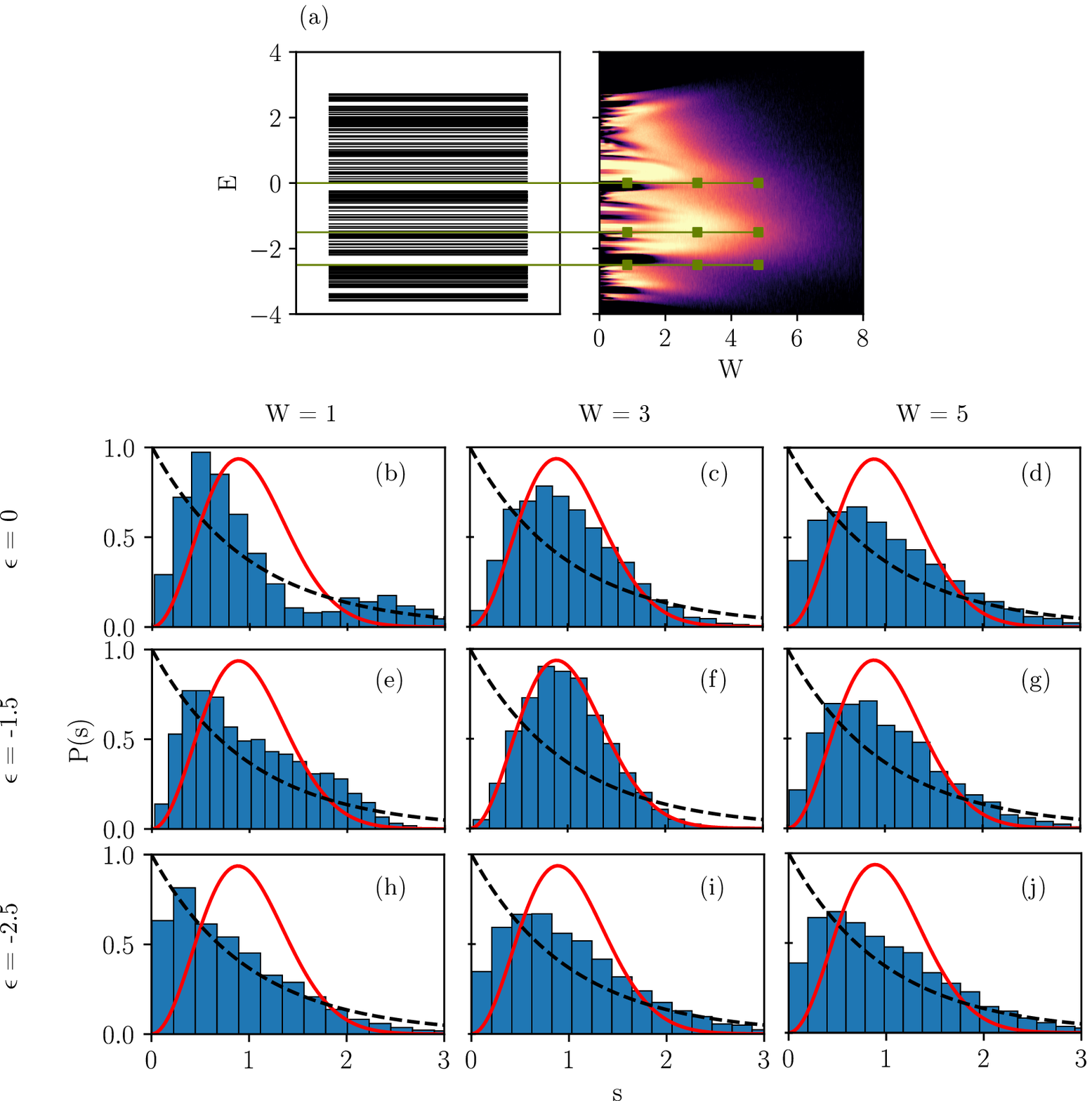}
\caption{(a) Energy levels for clean $n=5$ gasket together with phase diagram from \ref{fig:Chern_disord} at $\alpha = 1/4$. (b)--(j) distribution of the level spacings. Histograms are related to the spacings around (h, i, j) $\epsilon = -2.5$, (e, f, g) $\epsilon = -1.5$ and (b, c, d) $\epsilon = 0$. Calculations were performed for three disorder strengths $W = 1$ (b, e, h), $W = 3$ (c, f, i) and $W = 5$ (d, g, j)}
\label{fig:hist_SG}
\end{figure}

In Fig. ~\ref{fig:hist_SG} and \ref{fig:hist_SC} we show the level spacings distributions for three values of $\epsilon$ at $W = 1, 3, 5$ as examples. Numerical results are plotted together with Wigner-Dyson and Poisson statistics denoted by red solid and black dashed lines, respectively. We also present energy levels when the disorder is not introduced ((a) in Figs.~\ref{fig:hist_SG} and \ref{fig:hist_SC}). To follow the evolution of level spacings as a function of disorder strength $W$, we mark points on the phase diagrams with green squares, which correspond to the histograms below.

\begin{figure}
\centering
\includegraphics[width=0.7\columnwidth]{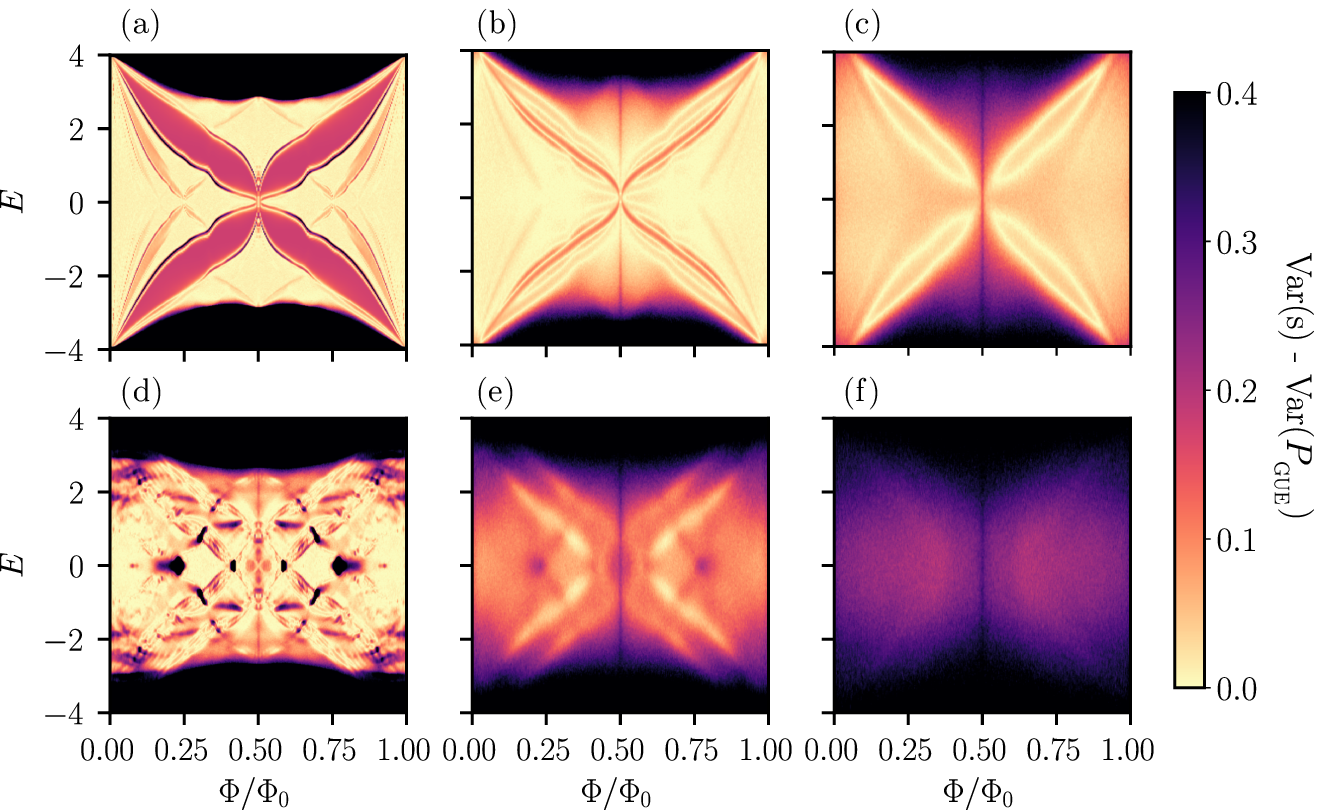}
\caption{Variance of the level spacings for a (a, b, c) square lattice and (d, e, f) Sierpi\'{n}ski carpet at (a, d) $W =1$, (b, e) $ W = 3$ and (c, f) $W = 5$ in the energy - flux plane.}
\label{fig:var_SC_square}
\end{figure}

\begin{figure}
\centering
\includegraphics[width=0.7\columnwidth]{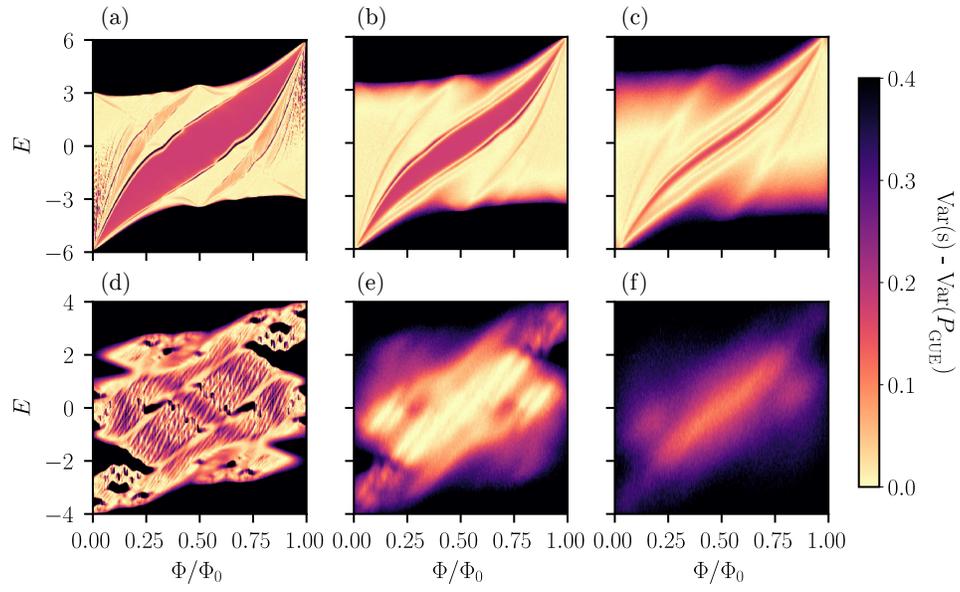}
\caption{Variance of the level spacings for (a, b, c) triangular lattice and (d, e, f) Sierpi\'{n}ski gasket at (a, d) $W =1$, (b, e) $ W = 3$ and (c, f) $W = 5$ in the energy - flux plane.}
\label{fig:var_SG_triangle}
\end{figure}

In Figs. \ref{fig:var_SC_square} and \ref{fig:var_SG_triangle} we present the $\textnormal{Var}(s) - \textnormal{Var} (P_{GUE})$ in the energy-flux plane at fixed $W = 1, 3, 5$ for fractal and regular lattices. In case of square and carpet, large variance exactly at $\alpha = 1/2$ is observed as systems are time-reversal invariant. For square and triangular lattices at small disorder, regions characterized by a large variance are separating delocalized states. This coincides with low DOS regions in the energy spectra.

\end{document}